\documentclass[12pt]{article}
\let\section=\subsection     \let\subsection=\subsubsection                %%
\usepackage{graphicx,epsfig}

\begin{document}
\begin{center}
   {\large \bf Nucleon resonances and the production of }\\[2mm]
   {\large \bf light vector mesons near thresholds\footnote{
   Supported by BMBF 06DR121 and GSI.}}\\[5mm]
   B.~K\"ampfer${}^a$, L.P.~Kaptari${}^{a,b,c}$, A.I.~Titov${}^{a,c,d}$\\[5mm]
   {\small \it 
   $^a$ Forschungszentrum Rossendorf, PF 510119, 01314 Dresden, Germany\\
   $^b$ Department of Physics, University of Perugia,
%   and INFN Sezione di Perugia, 
   via A. Pascoli, I-06100, Italy\\
   $^c$ Bogolyubov Laboratory of Theoretical Physics, JINR Dubna, 141980, Russia\footnote{
   permanent address}\\
   $^d$ Advanced Science Research Center, JAERI, %Japan Atomic Energy Research Institute, 
   Tokai, Ibaraki,  319-1195, Japan \\[8mm] }
\end{center}

\begin{abstract}\noindent
The production of the light vector mesons $V = \rho, \omega, \phi$
in the reactions $\pi N \to V N$ and $N N \to V N N$ near threshold
is studied. The subsequent electromagnetic decay $V \to \gamma^* \to e^+ e^-$
is particularly suited for exploring subthreshold $\omega N$ resonances.
\end{abstract}

\section{Introduction}

Baryon resonances represent an important part of the hadronic spectrum
which in turn reflects the pattern of excitations above the QCD ground state.
Their experimental investigation proceeds with various probes,
e.g., in photo- or electro-production processes at nucleons or nuclei.
Also in hadronic reactions the baryon resonances play an important role.
The theoretical investigation is based on coupled channel analyses, 
effective field theories, chiral perturbation theory, QCD sum rules etc.
The resonances once excited decay under emission of hadrons, real or virtual photons.
The decay channels we here focus on are such ones where light vector mesons
are emitted. An understanding of the elementary hadronic production processes
of vector mesons is of utmost importance for analyzing di-electron spectra
from intermediate heavy-ion reactions and hadron-nucleus reactions as well.
Corresponding experiments with the HADES detector \cite{HADES} started recently. 
This is the main motivation for the present note.

To illustrate some aspects of light vector meson production in hadronic
reactions let us mention three issues related to this topic.

(i) While for the processes $\pi p \to V N$ and $p p \to V p p$ 
(here $V = \rho, \omega, \phi$; $N$ stands either for the proton ($p$) 
or for the neutron ($n$))
some experimental data is at our disposal, one needs often a model to
deduce the corresponding reactions at the neutron. 
There are various examples for the ratio of the cross sections
$\sigma (p n \to V p n) / \sigma (p p \to V p p)$ which highlight
a strong energy dependence and clearly show that there is no simple
isospin factor relating these two channels.

(ii) There is a long list of various effects influenced by the hidden strangeness
content in the nucleon. Among them is the dropping mass of the $\phi$ meson
in nuclear matter, as predicted by the QCD sum rule approach \cite{QSR,Zschocke1}.
As shown in \cite{Zschocke2}, such an in-medium mass shift of the $\phi$
meson can considerably modify the di-electron spectrum in heavy-ion collisions.
This modification depends on the poorly known fraction of hidden strangeness
in the nucleon. Present lattice QCD calculations do not deliver a stringent
constrain of this fraction \cite{N_strangeness}. Otherwise, as the nucleon
is the core of visible matter in the universe, it is challenging to understand
its composition, and therefore any possible source of information is well come.
Our approach to this problem is to consider the sensitivity of $\phi$ meson 
production on a possible $\bar s s$ component in the nucleon wave function.

(iii) Below we are going to include, within an effective Lagrangian
model on tree level, resonances in the above mentioned hadronic reactions
and elaborate their specific importance for several observables.
Otherwise, we also demonstrate that certain reactions can be effectively
parameterized without inclusion of resonances.

The present note is organized as follows.
In section 2 we recapitulate the isoscalar-isovector ($\rho - \omega$)
interference near the $\omega$ threshold and emphasize that this effect can allow a
unique access to resonance properties.
Section 3 is devoted to a brief description of the combined
$\omega$ and $\phi$ production in $\pi N$ reactions with respect to the
OZI rule and its implications for the hidden strangeness content of the nucleon.
In section 4 we consider $\omega$ and $\phi$ production in $N N$ reactions.
Our conclusions can be found in section 5.

\section{Iscalar-Isovector Interferences in
$\pi N\to N e^+e^-$ Reactions as a Probe of Nucleon Resonance Dynamics}

\begin{figure}[ht] \label{fig_1} 
   \vskip -3mm
   \includegraphics[width=4.5cm,height=3.0cm,angle=-0]{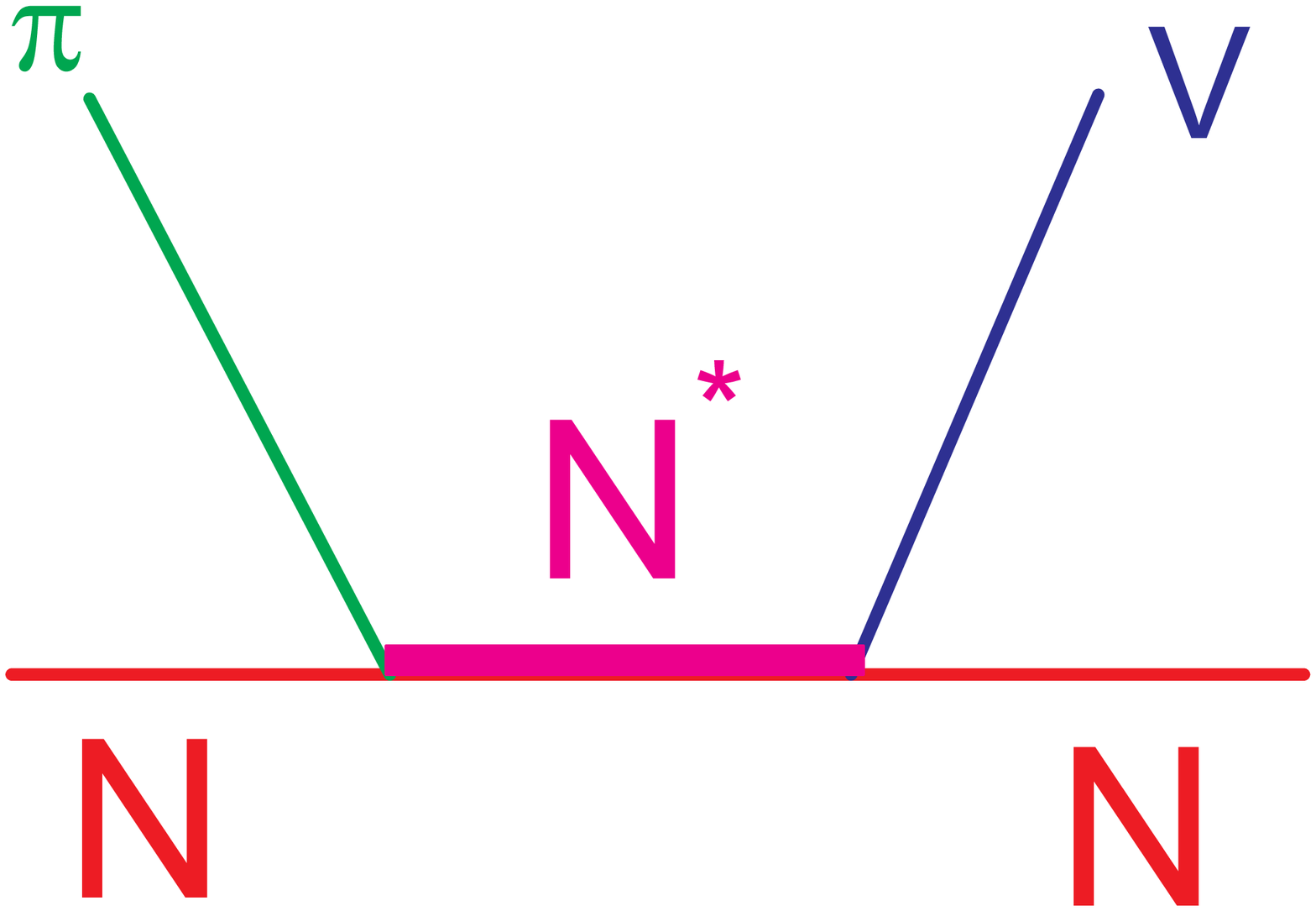}
   \includegraphics[width=4.5cm,height=3.0cm,angle=-0]{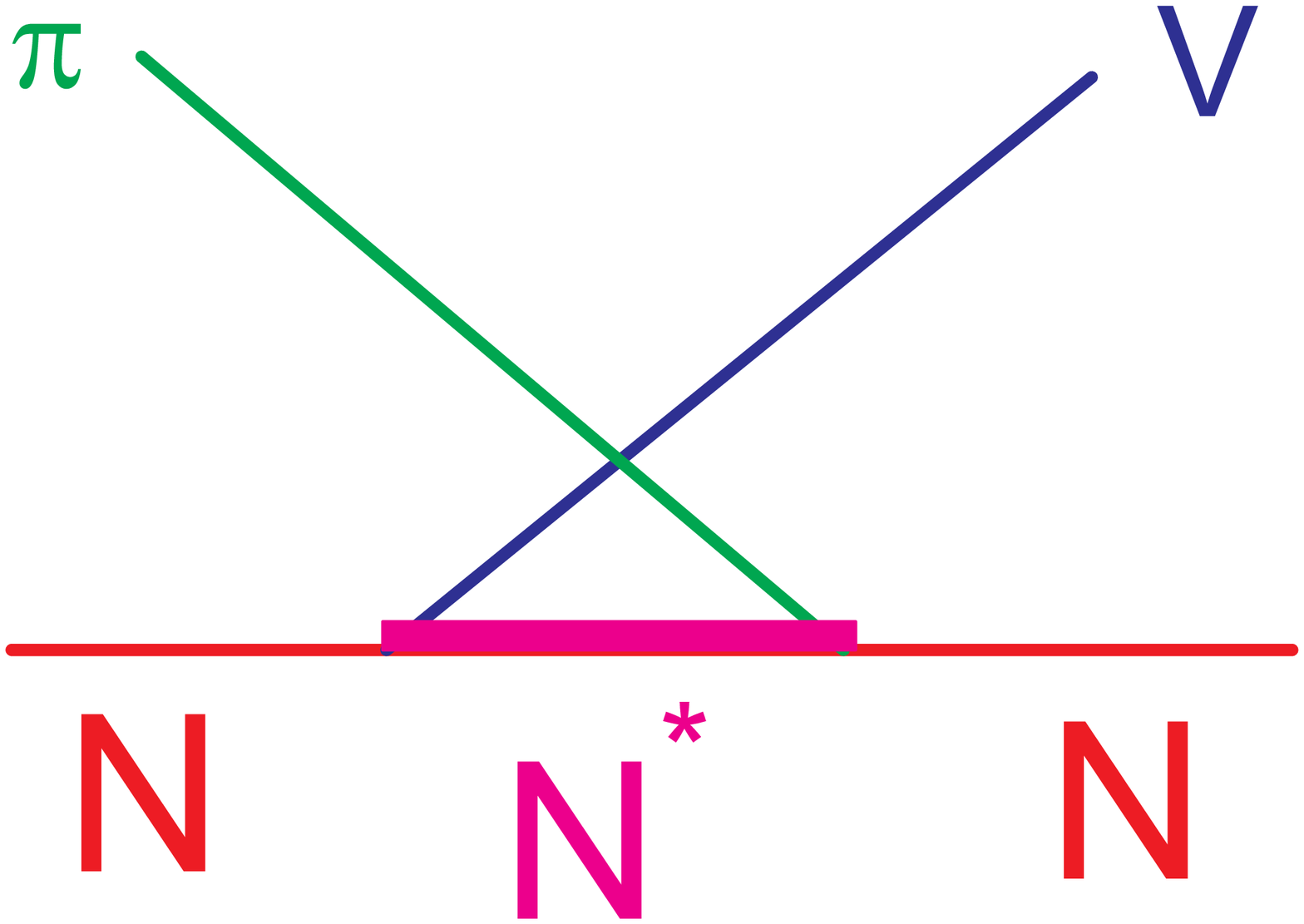}
   \includegraphics[width=4.5cm,height=3.0cm,angle=-0]{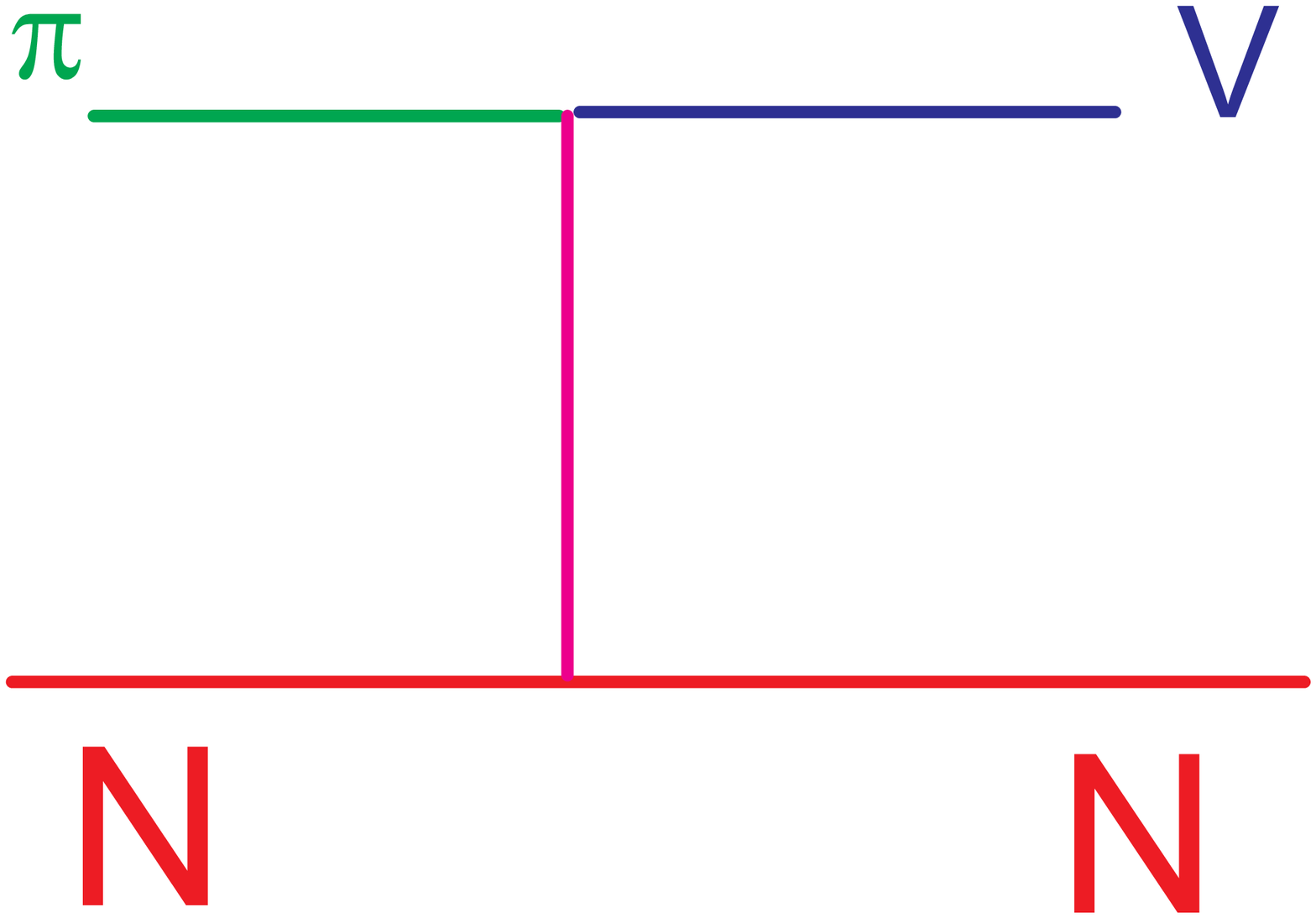}
   \vskip -3mm
   \parbox{14cm}
        \centerline{ 
        {\footnotesize Fig.~1:
        Tree level diagrams for the reaction $\pi N \to V N$.}}
\end{figure}

The tree level diagrams depicted in Fig.~1 serve as basis
for an exploratory study of the reaction $\pi N \to e^+e^- N$ with interfering
intermediate $\rho$ and $\omega$ mesons. This reaction, proposed in \cite{Madeleine1}
as a suitable probe of baryon resonance dynamics, is evaluated in \cite{we}
with resonances $N^*$ and $\Delta^*$ up to 1720 MeV without the t-channel
process displayed in the right part of Fig.~1; effective couplings are taken
from \cite{Riska_Brown} (cf.\ \cite{we} for details). Some illustrative results 
of this model exhibited in Fig.~2 agree in trends with the more
sophisticated approach of \cite{Madeleine2}: a destructive (constructive) 
$\rho - \omega$ interference in the reaction $\pi^- p \to e^+e^- n$ 
($\pi^+ n \to e^+e^- p$) below the $\omega$ threshold. Angular distributions
and the energy ($\sqrt{s}$) dependence of the invariant mass spectra can reveal
the role of individual resonances. A comparison of both isospin channels
can be used to disentangle the $N \omega$ subthreshold resonances.
Both reactions can be explored experimentally at GSI with the pion beams
and HADES in exclusive measurements. 
 
\begin{figure}[h] \label{fig_2}
   \includegraphics[width=6cm,height=4cm,angle=-0]{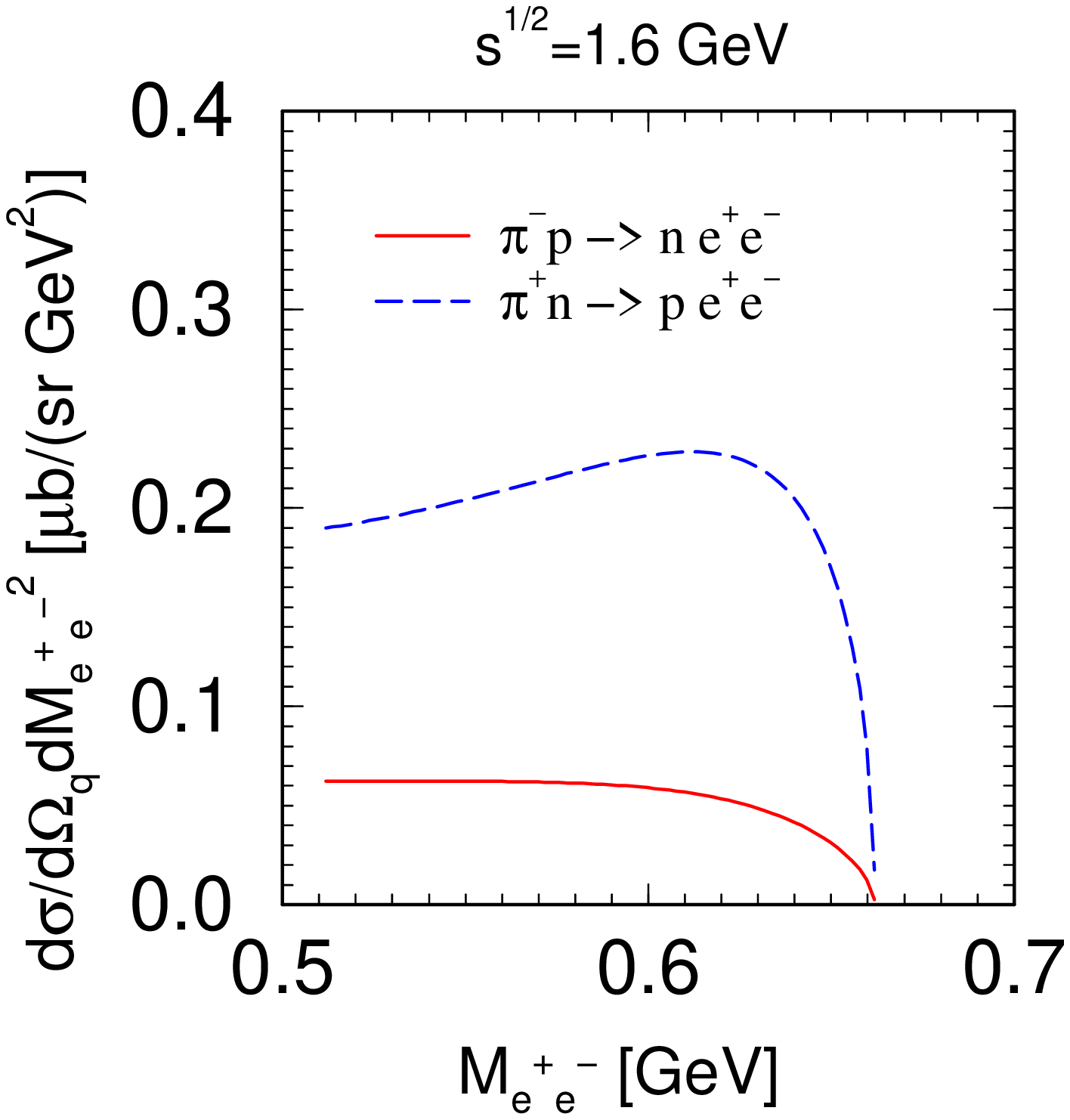} \hfill
   \includegraphics[width=6cm,height=4cm,angle=-0]{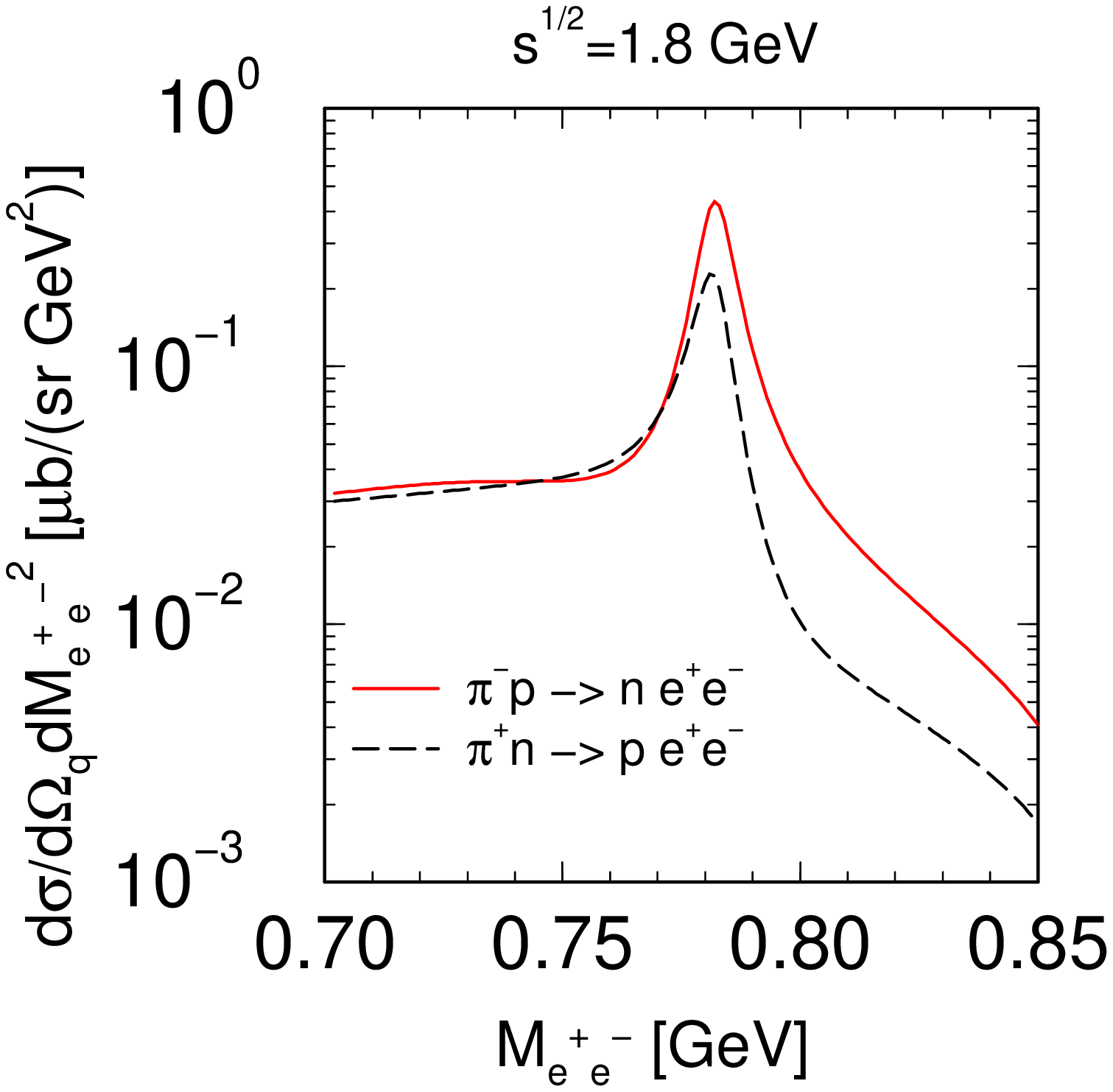}\\[2mm]
   \includegraphics[width=6cm,height=4.6cm,angle=-0]{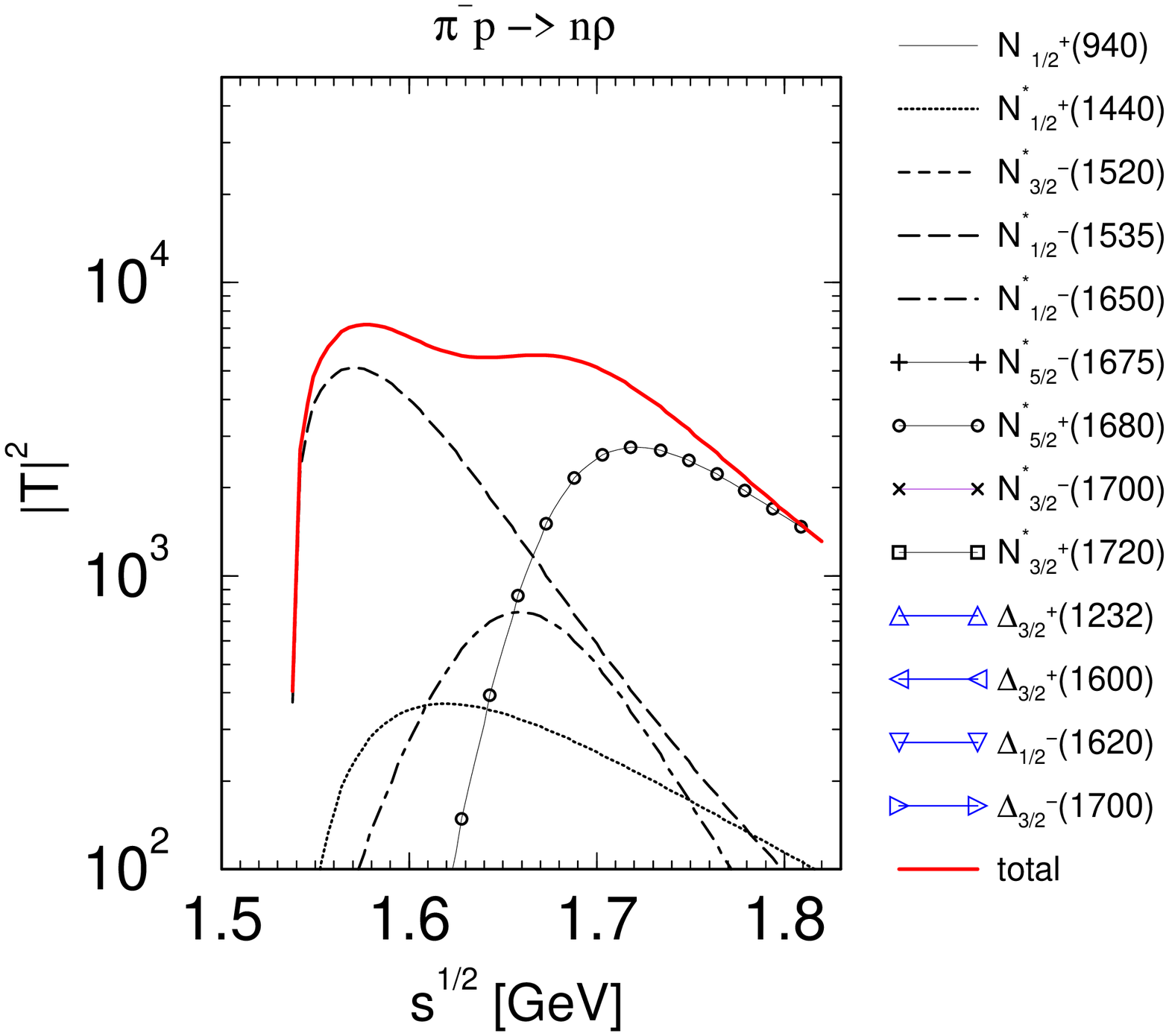} \hfill
   \includegraphics[width=6cm,height=4.6cm,angle=-0]{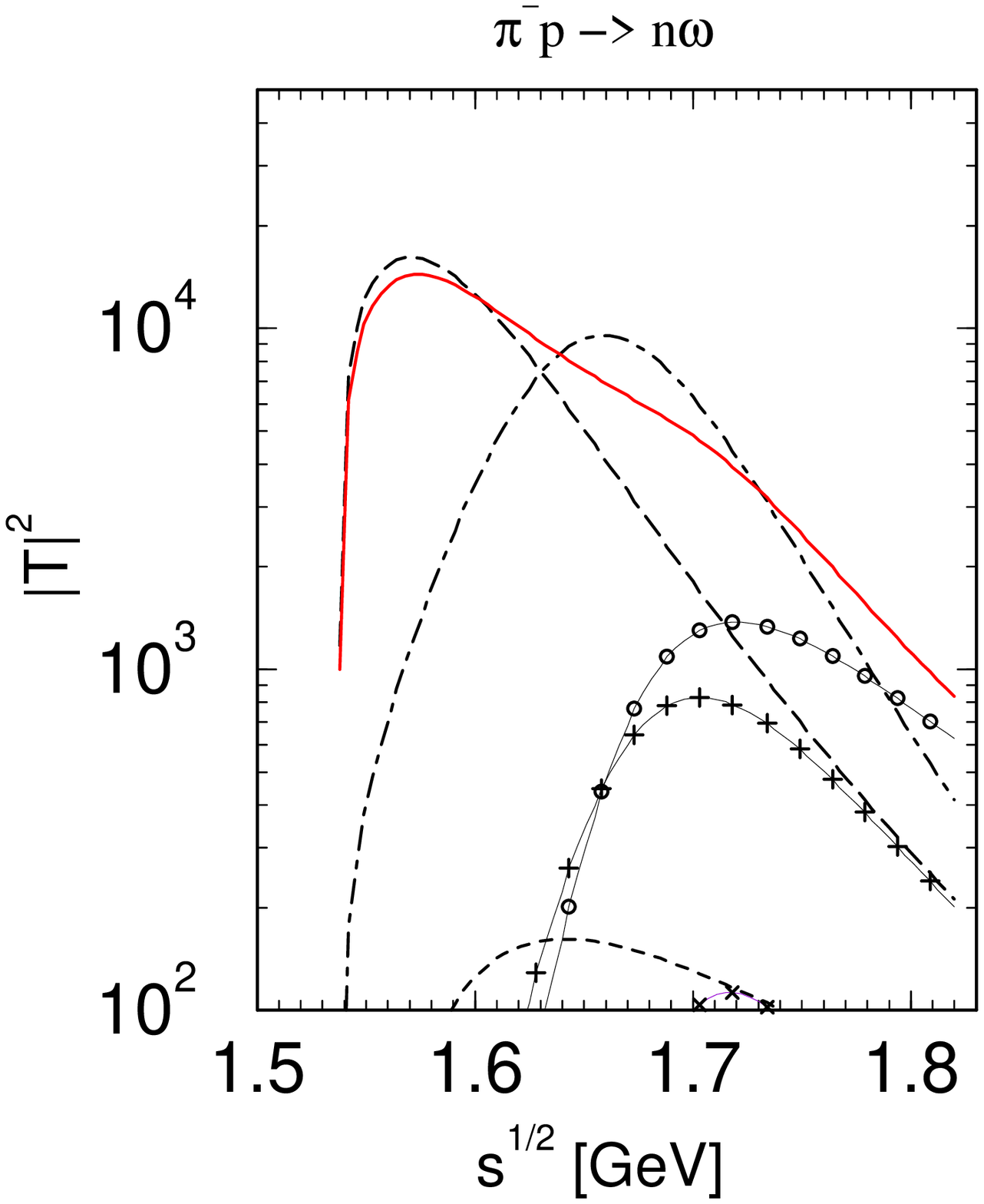}\\[2mm]
   \parbox{14cm}
        {\footnotesize Fig.~2: 
        Top: Differential cross section as a function of the $e^+ e^-$ invariant mass
        below (left) and above threshold (right) [the polar angle of the
        $e^+ e^-$ pair is $30^o$ in the $\pi N$ center-of-mass system].
        Bottom: Phase space averaged squared matrix elements for the individual
        resonances in the partial channels with an intermediate $\rho$ (left)
        and $\omega$ (right) meson as a function of the energy.}
\end{figure}

\section{Combined Analysis of $\omega$ and $\phi$ Production: OZI Rule}

A combined analysis of the $\omega$ and $\phi$ production in the reaction
$\pi N \to V N$ shows that the experimental data can be well described
by including s-, u- and t-channels. The contribution of the meson current
dominates \cite{we1}. Therefore, there is no need to introduce an anomalously
large coupling $g_{\phi N N}$ which could be interpreted, in the spirit
of \cite{Ellis}, as a hint to a substantial hidden strangeness fraction in the nucleon.
The standard OZI rule violation comes in this description from the
$\phi \rho \pi$ vertex, see \cite{Nakayama} for a discussion of this
issue.

Further investigations of photo-production of $\omega$ and $\phi$ mesons
at the nucleon \cite{Oh} do also not allow to deduce hints to
a sizeable admixture of hidden strangeness in the nucleon wave function. 

\section{The Reactions $N N \to \omega N N$ and $N N \to \phi N N$}

The tree level diagrams exhibited in Fig.~1 serve as building blocks
of diagrams for the corresponding processes $N N \to V N N$ in Fig.~3.
Therefore, one could expect that with adjusting the parameters of effective
Lagrangians to the reactions $\pi N \to V N$ a substantial part of the ambiguities
of in the parameter space is removed. However, additionally exchanged mesons and
form factors ruin this expectation. Rather, a combined analysis of both reactions
can reduce ambiguities, cf.\ \cite{pi_and_nucleon}.
\begin{figure}   \label{fig_3} 
   \includegraphics[width=4.5cm,height=3.0cm,angle=-0]{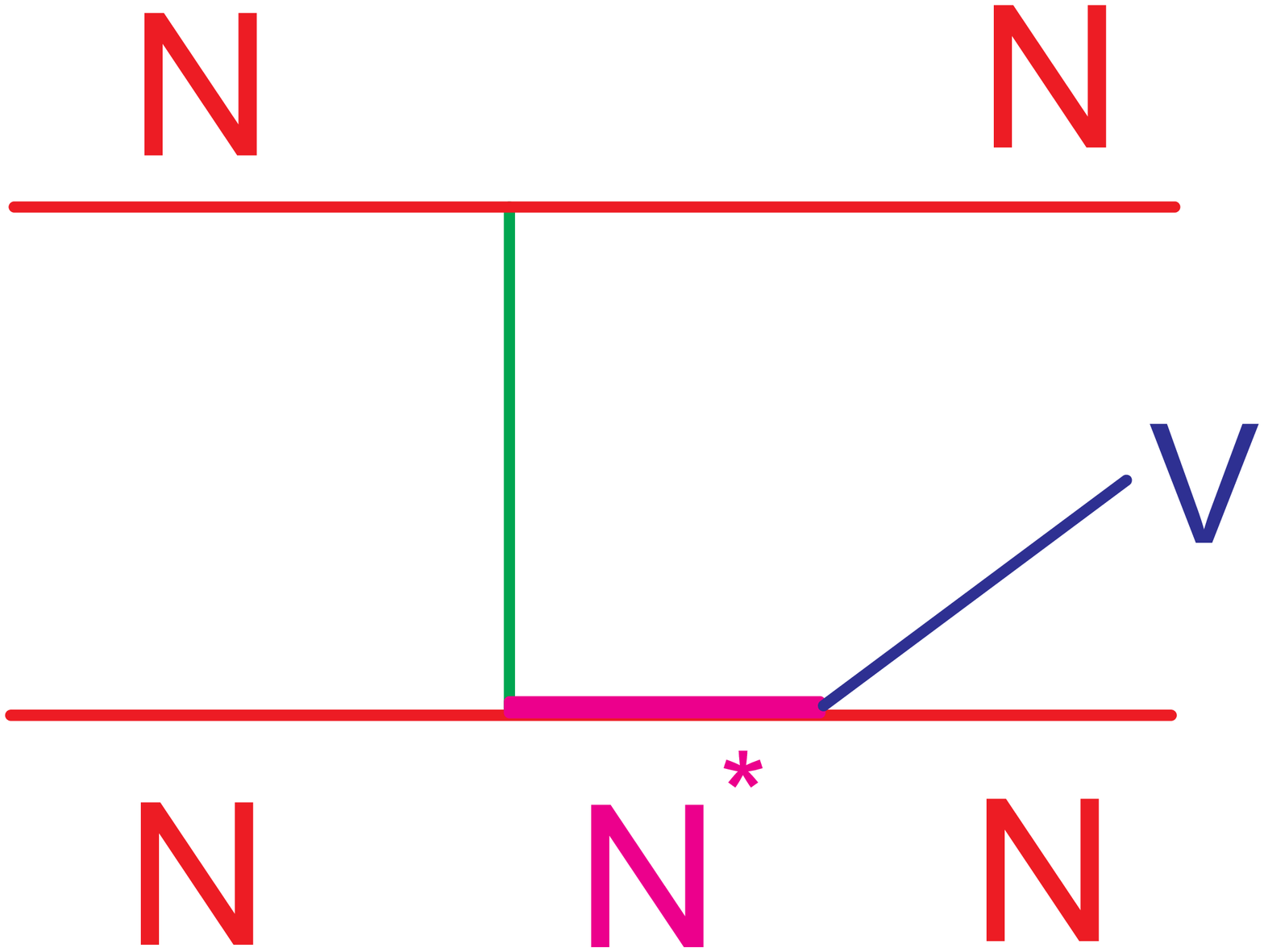}
   \includegraphics[width=4.5cm,height=3.0cm,angle=-0]{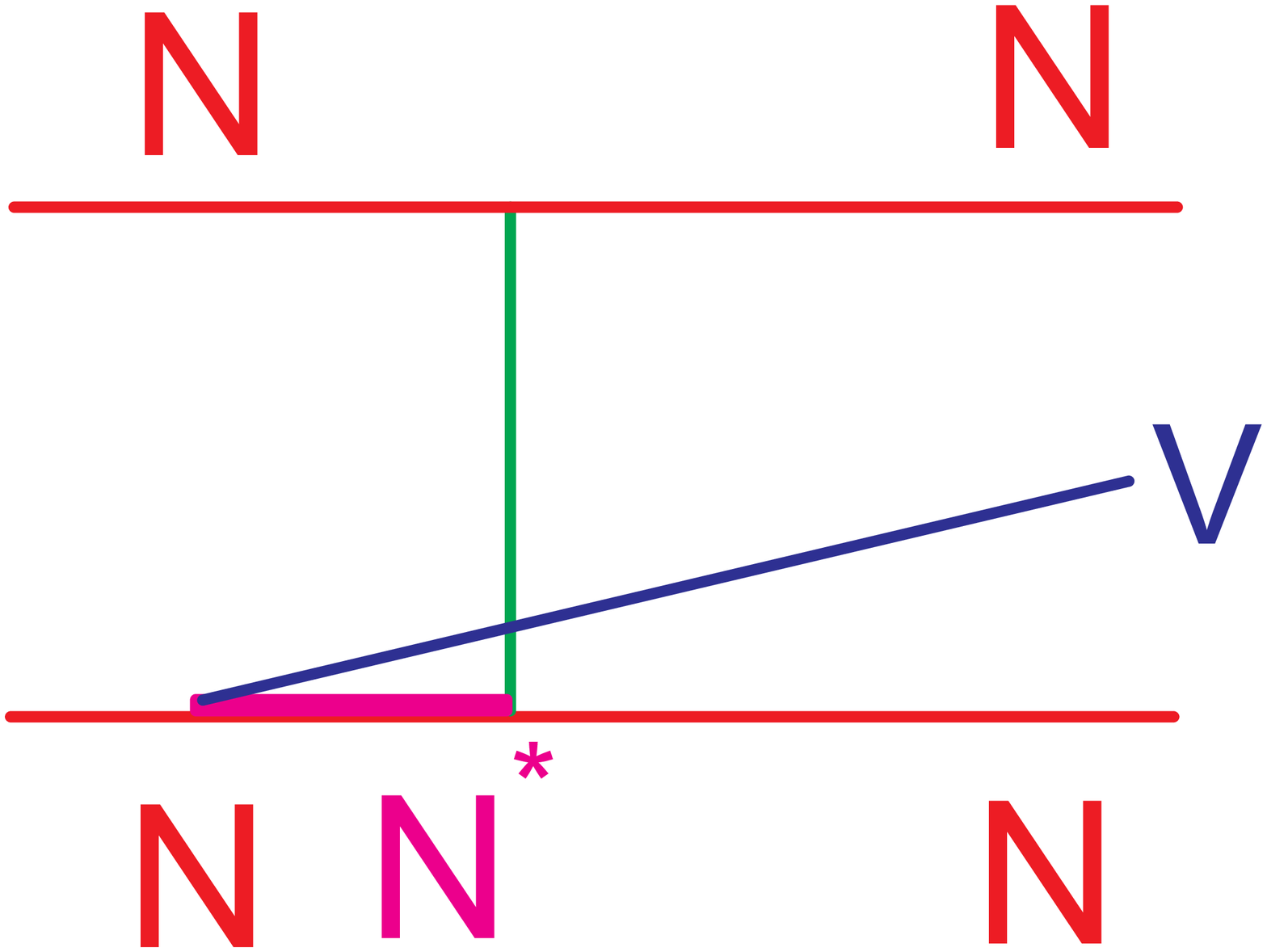}\\
   \includegraphics[width=4.5cm,height=3.0cm,angle=-0]{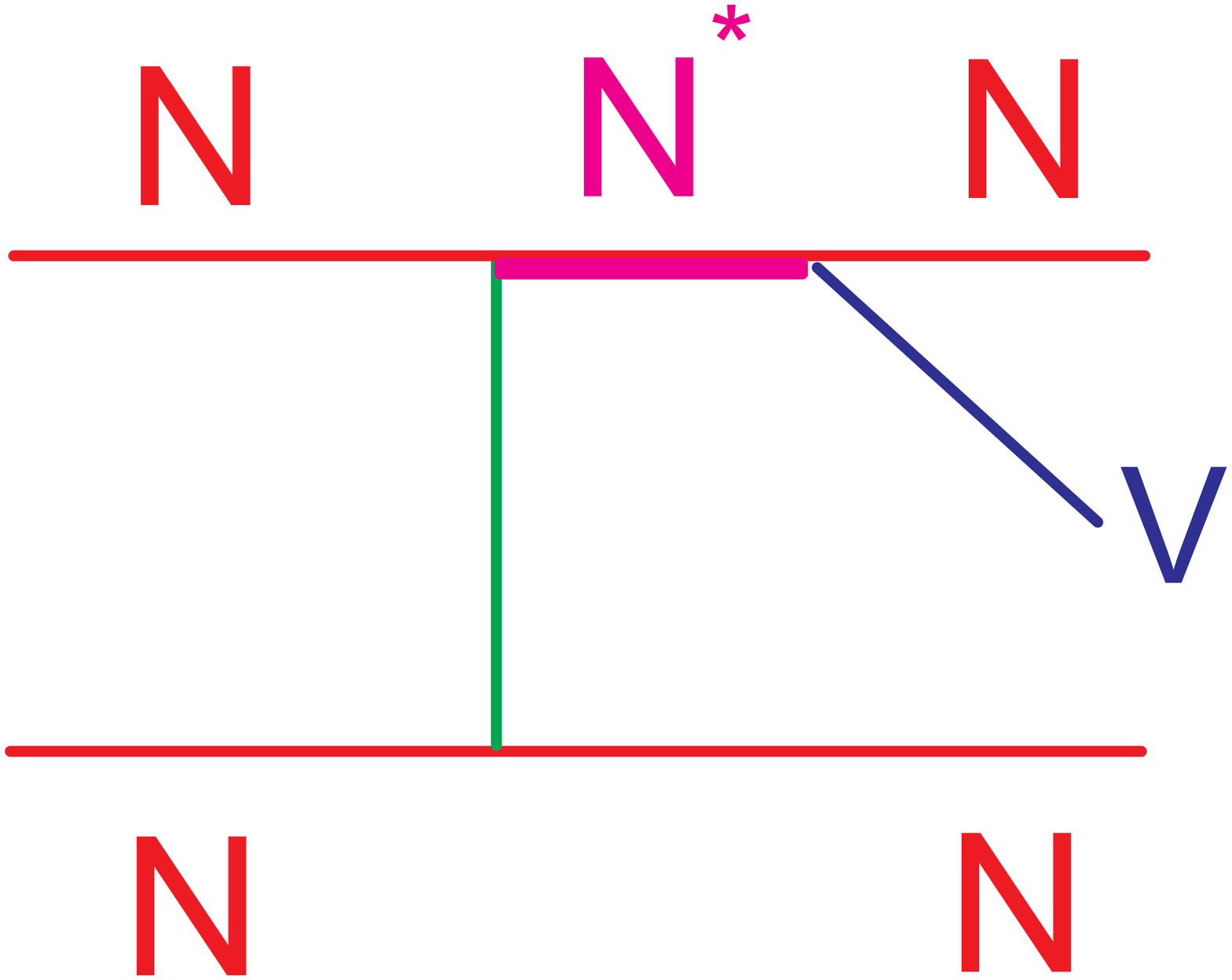}
   \includegraphics[width=4.5cm,height=3.0cm,angle=-0]{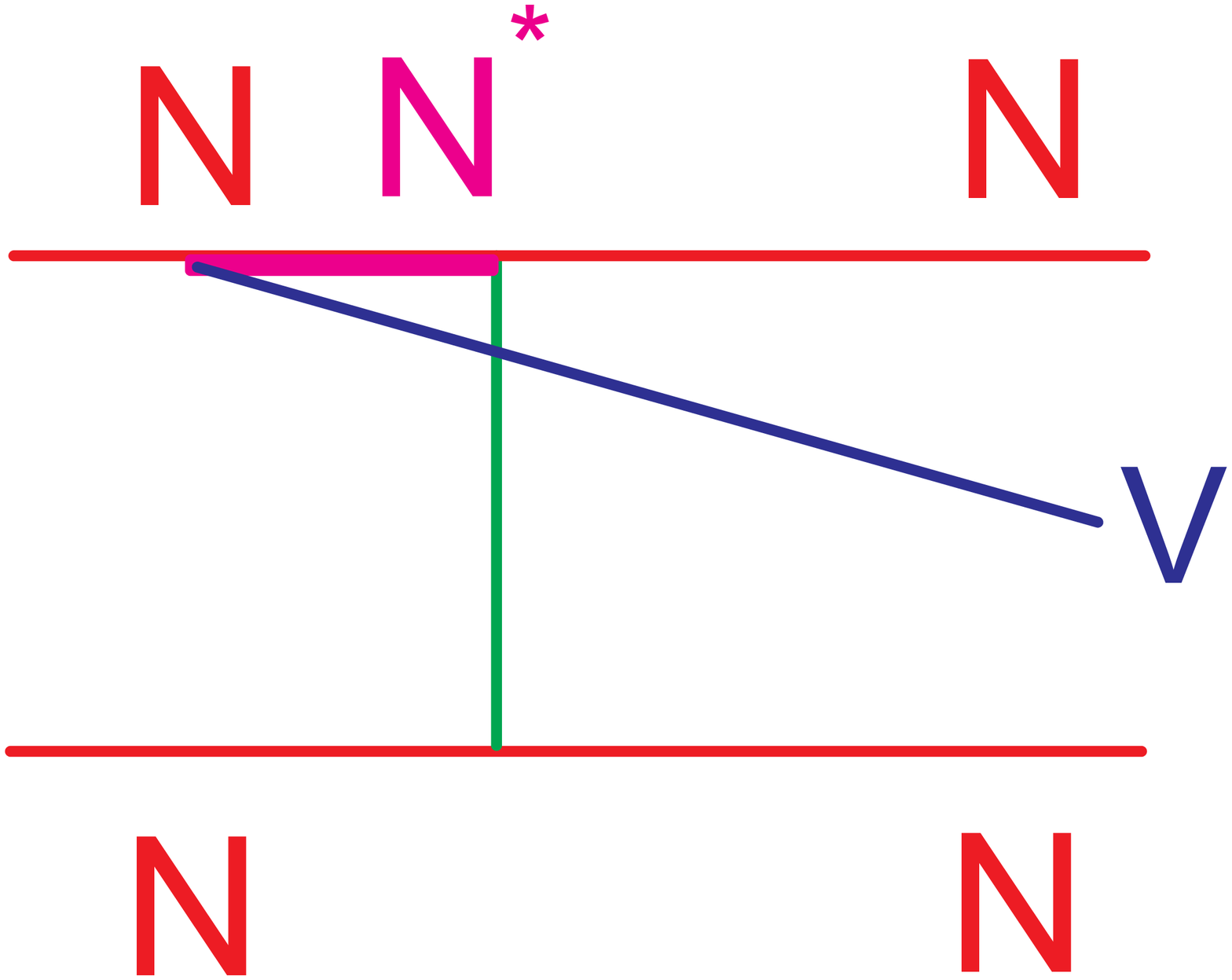}
   \includegraphics[width=4.5cm,height=3.0cm,angle=-0]{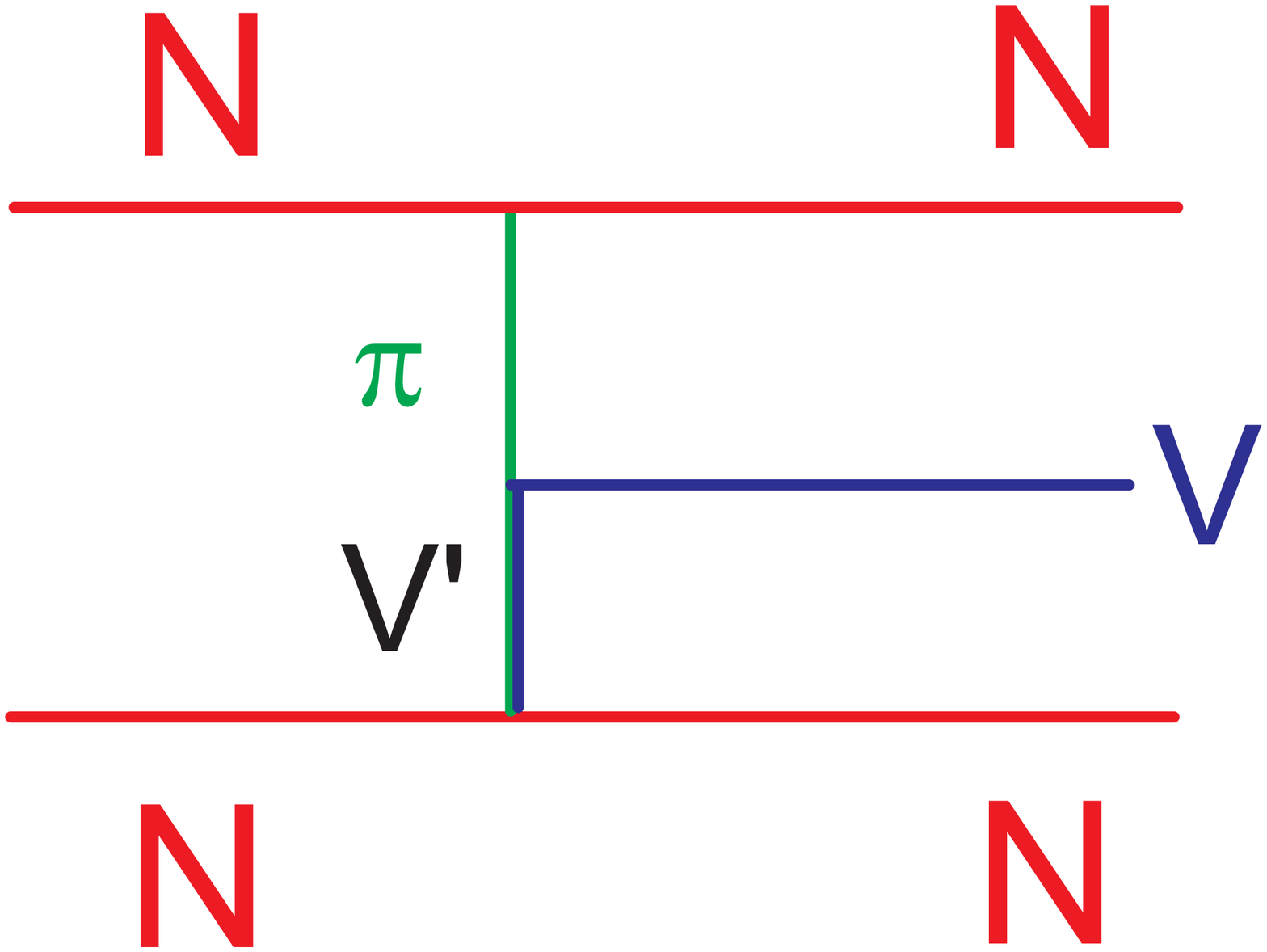}   
   \parbox{14cm}
        \centerline{
        {\footnotesize Fig.~3:
        Tree level diagrams for the reaction $N N \to V N N$.}}
\end{figure}

\subsection{$\phi$ Meson Production}

Fig.~4 exhibits a calculation \cite{pi_and_nucleon}
of the reaction $p p \to \phi p p$ within the described framework. 
To show the ambiguities inherent in the tree level effective Lagrangian 
approach two different parameter sets are used. One parameter set
delivers a dominating meson current for the sub-reaction 
$\pi^- p \to \phi n$ (Fig.~4 bottom left), while the other
employs a dominating nucleon current (Fig.~4 bottom right).
Correspondingly the interferences in the reaction $p p \to \phi p p$
are quite different (see Fig.~4 top). 

\begin{figure} \label{fig_4}
   \includegraphics[width=6cm,height=5cm,angle=-0]{CS_TETH_L80.epsi} \hfill
   \includegraphics[width=6cm,height=5cm,angle=-0]{CS_TETH_L80_.epsi}\\[1mm]
   \includegraphics[width=6cm,height=5cm,angle=-0]{pi_tot080.epsi} \hfill
   \includegraphics[width=6cm,height=5cm,angle=-0]{pi_tot080_.epsi}\\[3mm]
   \parbox{14cm}
        {\footnotesize Fig.~4: 
        Angular distribution of the cross section for the reaction
        $N N \to \phi N N$ at excess energy of 83 MeV (top, data from \cite{data1}) 
        and energy dependence of the total cross section for the reaction
        $\pi N \to \phi N$ (bottom, data from \cite{data2}). 
        Left (right) parameter set B (C) of \protect\cite{pi_and_nucleon}.
        Solid (dot-dashed/dashed) curves: total cross section 
        (contribution from the meson current with $\phi \rho \pi$ 
        vertex/contribution from the nucleon current).}
\end{figure}

The model for the reaction $p p \to \phi p p$ is used to deduce a parameterization
of the cross section cross section $p n \to \phi p n$ which serves as input
of a calculation \cite{tagging} of the reaction $p d \to d \phi p_{\rm spec}$,
where $p_{\rm spec}$ is a tagged spectator proton. A corresponding
measurement is feasible at COSY-ANKE \cite{ANKE1}.

\subsection{$\omega$ Meson Production}

The previous framework can be extended to study the production of $\omega$
mesons. A preliminary parameter adjustment to the angular distribution
at an excess energy of 170 MeV (Fig.~5 bottom right) describes the
energy dependence of the total cross section in fairly large range 
(Fig.~5 top left). (A similar approach is used in \cite{Tsushima}.
Note that present model does not yet include nucleon resonances.)
The isospin rotated reaction $p n \to \omega p n$ (Fig.~5 top right)
which results in a strongly energy dependent ratio of cross sections
(Fig.~5 bottom left) is presently used to study the reaction
$p d \to d \omega p_{\rm spec}$ for which already first results from
COSY-ANKE are at our disposal \cite{ANKE2}.

\begin{figure} \label{fig_5}
   \includegraphics[width=12cm,height=10cm,angle=-0]{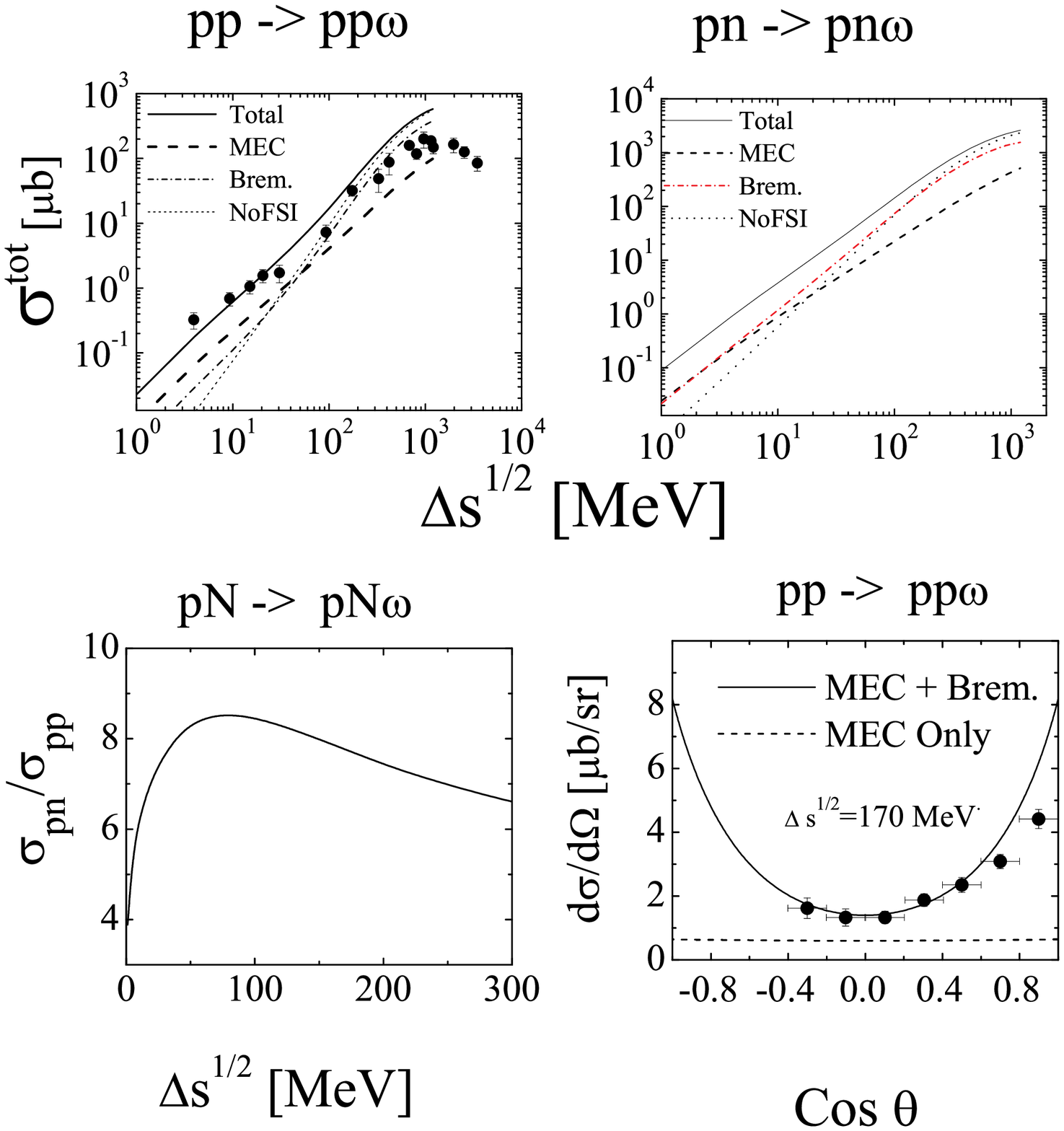}\\[6mm]
   \parbox{14cm}
        {\footnotesize Fig.~5: 
        Energy dependence of the total cross section for the reactions
        $p p \to \omega p p$ and $p n \to \omega p n$ (top, data compilation
        from \protect\cite{Tsushima}) and their ratio
        (left bottom). The parameters are adjusted to the angular distribution
        at excess energy of 170 MeV (right bottom, data from \protect\cite{COSY_TOF}).
        ''MEC'' denotes the contribution from the meson current (bottom right
        diagram in Fig.~3, while ''Brem'' depicts the
        contribution from the other diagrams. Final state interaction (FSI)
        among the nucleons is included along the lines of \protect\cite{we1}.}
\end{figure}

\section{Summary}

In summary we present a few selected examples of calculations of near-threshold
vector meson production in $\pi N$ and $N N$ reactions.
In doing so we employ tree level effective Lagrangians to accomplish simple
parameterizations of the elementary cross sections. The role of baryon resonances
in the reaction $\pi N \to e^+ e^- N$ is highlighted within a s-u-channel model
with coupling strengths deduced from a chiral quark model. A combined study of
of $\omega$ and $\phi$ meson production is mentioned not to reveal the need
of a noticeable $\bar s s$ shake-off the nucleon. The role of baryon resonances
for the reaction $p p \to \omega p p$ is presently under consideration.

\end{document}